\documentstyle[aas2pp4]{article}
\input epsf
\def\plotone#1{\centering \leavevmode
\epsfxsize= 0.8\columnwidth \epsfbox{#1}}

\def\plotonelocal#1{\plotone{#1}}
\def\plotonefull#1{}

\def\be{\begin{equation}}
\def\ee{\end{equation}}
\def\bea{\begin{eqnarray}}
\def\eea{\end{eqnarray}}

\def\muk{\mu{\rm K}}
\def\icm{${\rm cm}^{-1}$}

\def\cmm2{{\,\rm cm^{-2}}}
\def\cm2{{\,{\rm cm}^2}}
\def\cmm3{{\,{\rm cm}^{-3}}}
\def\gcmm3{{\,{\rm g\,cm^{-3}}}}

\def\fun#1#2{\lower3.6pt\vbox{\baselineskip0pt\lineskip.9pt
  \ialign{$\mathsurround=0pt#1\hfil##\hfil$\crcr#2\crcr\sim\crcr}}}

\def\cl{{\cal C}_l}

\def\C{{\cal C}}

\begin{document}
\twocolumn
\title{New CMB Power Spectrum Constraints from MSAM1}

G.~W.~Wilson\altaffilmark{1,2},
L.~Knox\altaffilmark{1},
S.~Dodelson\altaffilmark{3},
K.~Coble\altaffilmark{1},
E.~S.~Cheng\altaffilmark{2},
D.~A.~Cottingham\altaffilmark{4},
D.~J.~Fixsen\altaffilmark{5},
A.~B.~Goldin\altaffilmark{1},
C.~A.~Inman\altaffilmark{6},
M.~S.~Kowitt\altaffilmark{6},
S.~S.~Meyer\altaffilmark{1},
L.~A.~Page\altaffilmark{7}
J.~L.~Puchalla\altaffilmark{8},
J.~E.~Ruhl\altaffilmark{9},
and~R.~F.~Silverberg\altaffilmark{2}

\altaffiltext{1}{University of Chicago, 5640 S. Ellis Ave., Chicago, IL 
60637} 
\altaffiltext{2}{NASA/Goddard Space Flight Center, Laboratory for Astronomy
and Solar Physics, Code 685.0, Greenbelt, MD 20771}
\altaffiltext{3}{Fermilab, P.O. Box 500, Batavia, IL  60510}
\altaffiltext{4}{Global Science and Technology, Inc., NASA/GSFC Laboratory
for Astronomy and Solar Physics, Code 685.0, Greenbelt, MD 20771}
\altaffiltext{5}{Applied Research Corporation, NASA/GSFC Laboratory for
Astronomy and Solar Physics, Code 685.0, Greenbelt, MD 20771}
\altaffiltext{6}{Stanford Research Systems, Sunnyvale, CA  94089}
\altaffiltext{7}{Princeton University, Princeton, NJ  08544}
\altaffiltext{8}{Department of Physics and Astronomy, University of Pennsylvania, Philadelphia, PA  19104}
\altaffiltext{9}{University of California at Santa Barbara, Santa Barbara, CA 93106}
\setcounter{footnote}{0}
\vspace{.2in}

\begin{abstract}
We present new cosmic microwave background (CMB) anisotropy results
from the combined analysis of the three flights of the first Medium
Scale Anisotropy Measurement (MSAM1).  This balloon-borne bolometric
instrument measured about 10 square degrees of sky at
half-degree resolution in 4 frequency bands from 5.2 cm$^{-1}$ to 20
cm$^{-1}$ with a high signal-to-noise ratio. Here we present an overview
of our analysis methods, compare the results from the three flights,
derive new constraints on the CMB power spectrum from the combined
data and reduce the data to total power Wiener-filtered maps of the
CMB. A key feature of this new analysis is a determination of the
amplitude of CMB fluctuations at $\ell \sim 400$.  The analysis
technique is described in a companion paper (\cite{knox99}).
\end{abstract}


\section{Introduction}
The Medium Scale Anisotropy Measurement (MSAM) is a balloon-borne
telescope and bolometric radiometer designed to measure the anisotropy
in the cosmic microwave background (CMB) at angular scales near
0\fdg5.  The first two flights of MSAM1, reported in (\cite{cheng94})
(MSAM92) and (\cite{cheng95}) (MSAM94), observed overlapping fields on
the sky and demonstrated the repeatability of the measurement.  A
detailed comparison, showing consistency between these two flights,
was reported in (\cite{inman96}) and (\cite{knox98}).  A third flight
(\cite{cheng97}) (MSAM95) measured a nearby region of sky using the
same observing method.  This increased the experimental sky coverage
and sensitivity to the CMB anisotropy power spectrum.  A second
version of this instrument (MSAM II) with complementary frequency
coverage has since been flown.  This data set is still being analyzed.

\section{Instrument and Observations}

The MSAM1 instrument is described in (\cite{fixsen96a}).  We give a
summary here.  The actively pointed gondola is composed of a 1.4~m
off-axis Cassegrain telescope with a multimode bolometric radiometer.
A three-position chopping secondary throws the frequency independent
$\sim$0\fdg5  primary beam $\pm 0\fdg7$ tangent to the local horizon at 2~Hz.  The four spectral
channels at 5.7, 9.3, 16.5, and 22.5~\icm, each have bandwidth of
$\sim 1.5$~\icm.  The detectors' outputs are synchronously sampled at
32~Hz: 4 times for each of 4 positions of the secondary mirror, for a
total of 16 samples per chopper cycle.  Telescope pointing is
controlled with a star camera and gyroscope.  The configuration of the
gondola superstructure was changed between the 92 and 94 flights to
reduce possible reflection of ground radiation.  The improved
configuration remained for MSAM95.

\begin{figure}[bthp]
\plotonefull{/home/msam1/paper/pointing.eps}
\plotonelocal{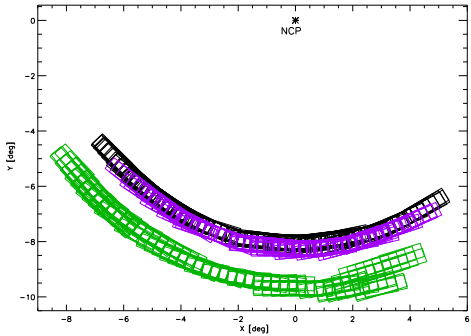}
\caption[pointing]{\baselineskip=10pt\small Locations in the data file
(flat) sky coordinates for the 1010 MSAM1 points.  The boxes show the relative
twist of the beam-pattern during the observation. The upper row of
points come from the overlapping flights, two years apart, of MSAM92
and MSAM94. The lower row are the MSAM95 points.}
\label{fig:point}
\end{figure}

All three flights were launched from the National Scientific Balloon 
Facility in Palestine, Texas.
The observing method, also described in (\cite{fixsen96a}), is a
slow azimuth scan of a region crossing the meridian above the North
Celestial Pole. For a period of 20 minutes the scan center tracks a
fixed spot on the sky as the earth rotates.  Afterwards, an
overlapping region is scanned. MSAM92 and MSAM94 observed at
declination $\delta=+82\fdg0$, MSAM95 at $\delta=+80\fdg5$. The
flights observed between right ascensions 14\fh2 and 19\fh5.  The lower
declination of the MSAM95 required a faster scan rate because of the
increased sky motion.  The sky coverage of all the MSAM1 flights is
shown in Fig.~\ref{fig:point}.

\section{Data Reduction}
The data from each of the three flights of MSAM1 are independently
reduced in the same manner.  A detailed discussion of the analysis for
each of the three flights is available in
(\cite{cheng94,cheng95,cheng97}) respectively.  We outline the process
here.

1) Spikes caused by cosmic rays are removed from the time stream by a
filtering and peak detecting technique which results in the deletion
of 5\% to 10\% of the data.  Samples are also lost due to spurious
electrical pickup and telemetry dropouts.  For each of these cuts, a
full chopper cycle is deleted.  The total loss is between 10 and 30\%
of the raw data.

2) The detector time streams are demodulated in two ways -- each
resulting in an independent instrumental beam pattern and
corresponding instrumental window function.  If $T_L$, $T_C$, and
$T_R$ are the sky temperature at the left, center, and right position
of the beam during a chopper cycle, the single difference demodulation
is $T_R - T_L$, making an antisymmetric beam pattern, while the double
difference is $T_C - (T_L + T_R) / 2$, making a symmetric
beam-pattern. Optimum weighting for the demodulations are determined
from Jupiter observations.  The instrument noise is uncorrelated
between between the two demodulations.

3) The data are calibrated using scan and raster observations of
Jupiter. The brightness temperature of Jupiter is reported in
(\cite{goldin97}). Of the two models presented in that paper, we use
the temperatures based on the ``Rudy'' model (\cite{rudy87b}).  The
error in the calibration is estimated to be 5\%, dominated by the
uncertainty in the Jupiter temperature.

4) The Jupiter raster observations, performed during each flight, are
the basis for the high fidelity determination of the beam pattern for
each demodulation.  Beam-pattern uncertainties are dominated by
arcminute-scale pointing uncertainties.

5) The estimate of the instrument noise is determined from the
variance in 100~s segments of the demodulated data after the removal
of a slowly drifting offset.  The offset ranges from 1 to 6~mK in
MSAM94 and MSAM95 with an offset of 10~mK R-J in all channels in
MSAM92.  The drift in the offset is small compared to its value.
Because the removal of the offset correlates the noise on time scales
longer than the detector time constant, the remainder of the data
reduction incorporates the {\it full} noise covariance matrix.

6) The data are binned according to both the position on the sky and
the twist of the demodulated beam pattern during each complete chopper cycle.  The bin
size for the twist dimension is determined by defining a ``Binning
Degradation Factor,'' $BDF = \sqrt{\sigma^2 +
\left<\delta^2\right>}/\sigma$ where $\sigma$ is the estimated
instrument noise and \newline
\hbox{$\left< \delta^2 \right> = \int\left(
|B_1(\vec{k}) - B_2(A\vec{k}) |^2 C_{\rm cdm}(\vec{k}) \right)d\vec{k}$}
is an estimate of the expected error in the estimate of the signal due to the
twist bin size. $\left< \delta^2 \right>$ is determined using the
standard cold dark matter correlation function $C_{\rm cdm}(\vec{k})$ convolved
with beam-patterns, $B_i(\vec{k})$ twisted with respect to each other by
the rotation matrix $A$.  A similar construction is used to define the
$BDF$ for the spatial binning.  The $BDF$ can be thought of as the
factor by which the sensitivity of the data set is decreased due to the
choice of bin size.  The bin sizes are chosen to hold the $BDF$ to
values less than 1.1. This results in 5\arcdeg twist bins and 
14\arcmin bins in sky position.

7) The calibrated data are analyzed to provide measurements of
brightness in the four spectral channels as a function of bin.  The
linear combination of the spectral channels which minimizes
the sensitivity of galactic dust foreground and matches the signature
of a CMB thermal fluctuation over the spectral range of the instrument
channels is found and an estimate of CMB anisotropy and dust optical
depth for each bin is produced. This is done by fitting the data for each
bin in the four channels to a two parameter model of sky and dust.

\section{Comparison of MSAM92 and MSAM94}

The overlapping regions of MSAM92 and MSAM94 (see
Fig.~\ref{fig:point}) are used to compare their estimated sky
signals. This can place a limit on how much of the signal could be
attributed to instrumental artifacts or other local contamination.
While straightforward in principle, a simple comparison is not
possible despite the large degree of overlap. The beam centers for
each sample do not line up perfectly and because of the twist
dimension in the binning, there are few bins that are populated in
both flights. In (\cite{inman96}) the bin size was expanded over the
criterion in the previous section and those bins with sufficient data
were differenced.  With rather reduced sky coverage, (\cite{inman96})
found no signal in the differenced data.

\begin{figure}[bthp]
\plotonefull{/home/knox/proj/msam/f92_shade9294.eps}
\plotonelocal{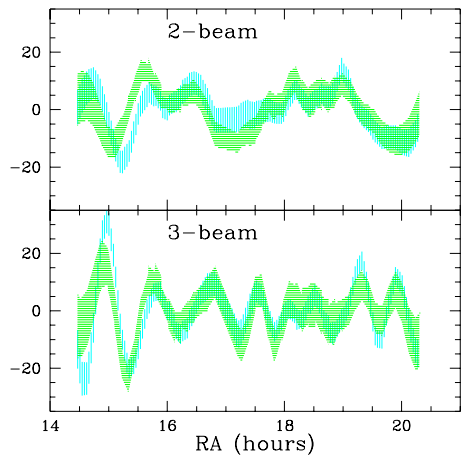}
\caption[comparison]{\baselineskip=10pt\small 
Most likely signal in 1992 data set, given the COBE-normalized
standard CDM power spectrum and the 1992 data (vertical lines) or
the 1994 data (horizontal lines).  The shaded area is the 68\%
confidence region.  Single-difference (or ``2-beam'') data in top panel, 
double-difference (or ``3-beam'') in the bottom panel. 
}
\label{fig:compare}
\end{figure}

An alternative procedure for comparing the two measurements has been
previously reported in (\cite{knox98}).  Here, an assumed power
spectrum for the CMB fluctuations is used to make a prediction of the
most likely signal in the 1992 data set, given the 1992 pointing
information.  This is compared to the most likely value of the signal
in the same 1992 data set but given the 1994 pointings.  This ``most
likely'' signal is determined by applying a Wiener filter to the
data.  See (\cite{knox98}) for details.  In Fig.~\ref{fig:compare} we
see that the two data sets predict very similar signals for the 1992
data set using either the 1992 or 1994 data.

We quantify ``very similar'' by use of the likelihood ratio statistic.
The two hypotheses are 1) the signals are correlated as one
would expect (given the two sampling strategies and an assumed power
spectrum) and 2) the signals are uncorrelated between data sets.
We use the natural log of the likelihood ratio statistic,
which is a quadratic operator on the data denoted by $\beta$
\footnote{Also see (\cite{Tegmark98}) on the optimization of quadratic
comparison statistics.}.  For the 1992 and 1994 data sets
(\cite{knox98}) has found $\beta = 12.8$, which means that hypothesis
1 is $e^{12.8}$ times more likely than hypothesis 2.  A frequentist
interpretation of $\beta$ is given by calculating the expected mean and
standard deviation of the statistic under the different hypotheses. The
result is $15.0 \pm 4.1$ (hypothesis 1) and $-58.4 \pm 27.4$
(hypothesis 2). This analysis is in agreement with (\cite{inman96})
that it is extremely unlikely that the data sets are caused by a signal
that is uncorrelated between experiments. Based on these analyses, we conclude
that the signal comes from the sky and not from the instrument or
local environment.

\section{Likelihood Analysis}
The data set from the three flights of MSAM has been reduced to 505
measurements of the CMB sky, for each of the two demodulations.
We model this data, $d$ as due to signal and noise
\be
d_i = s_i +n_i
\ee
where $i$ runs from 1 to 505 over the single-difference demodulation
and from 506 to 1010 over the double-difference demodulation, and
the signal, $s$,  is related to the true temperature field, $T$,
by
\be
s_i= \int_{\Omega}B(\vec{x}-\vec{x}_i)T(\vec{x})d\vec{x}.
\ee
Here, $B(\vec{x})$ is the (single or double
difference) beammap, and $\vec{x}_i$ specifies the pointing.
We assume that both the signal and noise are Gaussian-distributed
with zero mean with covariance matrices which we denote by
$S_{ij} = \langle s_is_j\rangle$ and $N_{ij} = \langle n_i n_j \rangle$.

The noise covariance matrix, $N$, is block-diagonal with
each block representing the noise correlations of a single
demodulation from a single flight.  The noise covariance matrix is
singular due to the independent offset removals from each of the three
flights (two each in MSAM92 and MSAM94).  This constraint must be
explicitly projected out of the data which we do with an SVD inversion of
$N$.

The signal covariance matrix is linearly related to the angular
power spectrum of
the temperature field, $C_l$.  The likelihood of this power spectrum,
given the data, noise matrix and our assumptions of Gaussianity is
\be
\label{eq:like}
L(C_l) = \frac{e^{-\frac{1}{2}\vec{d}C^{-1}\vec{d}^T}}
{(2\pi)^{N/2}\sqrt{\det{C}}}
\ee
where $\vec{d}$ is the $n$-element vector of observations and $C=[S(C_l)
+ N]$ is the $(n\times n)$ covariance matrix of the observations. 
We use this likelihood to place limits on the
power spectrum of fluctuations. 

For this analysis we parameterize the theoretical signal covariance
matrix, $S$, with the power spectrum, $\cl \equiv l(l+1) C_l/(2\pi)$,  
broken into bands denoted by $B$ so that
\be
\label{eq:cl}
\cl = \sum_B \chi_{_{B(l)}} \C_B,
\ee
where $\C_B$ denotes
is a flat power spectrum within band $B$ with amplitude $C_B$.  That is,
\be
\C_B = \frac{l(l+1)}{2\pi}C_B.
\ee
The sum runs over the bands in $l$-space with 

\be
\chi_{_{B(l)}}  = \left\{ 
\begin{array}{r@{\quad:\quad}l}
1 & l_<(B) < l < l_>(B) \\
0 & {\rm otherwise}
\end{array} 
\right . 
\ee
This parameterization of $\cl$ is completely general and its usefulness  will
become apparent below.

The calculation of the likelihood requires the inversion of the
$(n\times n)$ covariance matrix $C$.  It has been shown in
(\cite{bond94a,tegmark96b,bunn97,bond96a}) that a substantial reduction in the rank
of $C$ can be achieved by working in the signal-to-noise eigenmode
basis. This is true even in a high signal-to-noise case like that of
MSAM1. For this data set, we achieve a compression by a factor of 1.8
in the rank of $C$ by ignoring modes with signal-to-noise ratio of
less than 0.03.  Working in the signal-to-noise eigenmode basis has
the added benefit of automatically projecting out the eigenmodes
associated with the offset removal.  Thus, only one initial SVD of the
covariance matrix is required (to zero the infinite eigenvalues).
Inversions of the covariance matrix in the signal-to-noise eigenmode
basis are then done using faster methods such as Cholesky
decomposition.

\section{The Flat Band-Power}

As has been done previously for each of the data sets individually
(\cite{cheng94,cheng95,cheng97}),  
we calculate ``flat band-powers'' for each demodulation.
That is, we assume the entire power spectrum is flat with amplitude
$\C_l=\C_B$ and calculate the likelihood of this amplitude.    
Table~\ref{table:bp} gives the flat band-powers (maximum likelihood 
values of $\sqrt{\C_B}$)
for the three flights of MSAM1 for the single and double
difference demodulations. The error bars indicate where the likelihood
falls to $e^{-1/2}$ of the maximum.

\begin{table}[htbp]
\centering
\vspace{1ex}
\begin{tabular}{c||c|c}
Flight          &       Single Diff.    &  Double Diff. \\ \hline\hline
MSAM92  & $48 \pm 11$   & $54 \pm 10$  \\  \hline
MSAM94  & $35 \pm 6$    & $45 \pm 9$    \\      \hline
MSAM95  & $51 \pm 7$    & $56 \pm 7$    \\      \hline
all three & $47 \pm 5$ & $53 \pm 5$     \\ 
\end{tabular}
\caption{Flat Band-power Estimates for MSAM1 in $\muk$}
\label{table:bp}
\end{table}

\section{Radical Compression}

Flat band-powers, together with the diagonal parts of the 
window function matrix, often simply called the window function, have
traditionally been the main results of CMB experiments.  When taken
together, they are the raw ingredients for constraining
the power spectrum and cosmological parameters.

The parameters, $a_p$, are found by minimizing the $\chi^2$ where
\be
\chi^2 = \sum_B (\sum_l f_{Bl} \cl(a_p) - \C_B)^2/\sigma_B^2,
\label{eq:chisq}
\ee
where $B$ runs over different data sets, $\C_B$ and $\sigma_B$ are 
the band-powers and their standard
errors respectively, and $f_{Bl}$ is a filter which, when summed
over the power spectrum, $\C_l$, gives the theoretical prediction
for the band-power, $\C_B$.  The filter is usually 
constructed from the familiar window function, $W_{Bl}$
\be
\label{eq:iwf2f}
f_{Bl} = {W_{Bl}/l\over (\sum W_{Bl}/l)}.
\ee
We call the filter given by this equation the window function filter.
The parameters $a_p$ could be cosmological parameters (e.g., $\Omega_b$,
$\Omega_{\Lambda}$, $H_0$, etc.) or parameters from a phenomenological
power spectrum.  

\subsection{Problems With Flat Band Powers ...}

Using flat band-powers as a form of radical compression  has
the following drawbacks:

\begin{enumerate}

\item The actual sky power spectrum is not flat. \label{notflat}

\item The expectation value of the band power is not given by
summing the window function filter over the power spectrum, and thus
the window function filter should not be used in Eq.~\ref{eq:chisq}.
The expectation value is only given by this sum
in the limit that the data points have no signal correlations.
\label{nooffdiag}

\item The method provides no estimate of the correlation 
between the errors in the estimates of
$\C_B$ from different demodulations. \label{nocorre}

\item The constraints on the parameters are not Gaussian, even though this
assumption is implicit in the $\chi^2$ minimization. \label{nogauss}

\end{enumerate}

Problems \ref{notflat} and \ref{nocorre} are well known deficiencies
of the bandpower approach.  Problem \ref{nogauss} has been emphasized
in (\cite{bond98b}), where an approximate solution was given.  
Here we focus on problem \ref{nooffdiag}, which
has been discussed previously in (\cite{bond98a}).  We illustrate the
potential severity of the problem with an extreme example.  Consider a
total power mapping experiment with angular resolution of FWHM =
$30$\arcmin\ which has measured a $5\arcdeg\times 5\arcdeg$ patch of
the sky.  The window function 
filter for this experiment is $f_l \propto W_l/l =
\exp(-l^2\sigma_b^2)/l$ where $\sigma_b \simeq {\rm FWHM}/2.355$.  
Note that this filter peaks at $l=0$ indicating that the experiment is most 
sensitive to flucutations on very large angular
scales. However, the data set is not actually sensitive to the lowest
spatial frequencies at all.  The problem lies in having ignored the
off-diagonal terms.  The filter function actually makes sense only if
the data points are all far apart on the sky so that $S_{ij}$ is
diagonal.  For the example given, correlations between the points on
the sky are making the data set insensitive to fluctuations on large
scales.  Because using the diagonal component of the window function to
define the filter ignores these correlations, we get a
nonsensical result.  For most actual data sets, the problem is not
quite so severe but this example illustrates the potential pitfall.

\subsection{... and Solutions}
\label{sec:estimate}

Solutions have been found to all of these problems
(\cite{bond98a,bond98b,knox99}).  Here we briefly review them.

Problem~\ref{notflat} can be solved by breaking the
power spectrum into several bands as in Eq.~\ref{eq:cl}, and then finding
the amplitudes of these bands, $\C_B$, that maximize the likelihood.
We find this maximum by iterative application of a quadratic estimator,
as has been done for COBE/DMR (\cite{bennett96}) and 
Saskatoon (\cite{netterfield96}) data in (\cite{bond98a})
and on simulated MAP data in (\cite{oh99}).
By calculating the covariance matrix of the set of $\C_B$ we also
solve problem~\ref{nocorre}.  

Because physical power spectra are not actually flat across these
bands, we need a means of taking a general power spectrum, $\C_l$,
and turning it into a prediction for $\C_B$.  In other
words, we need to be able to calculate the expectation value of
$\C_B$, $\langle \C_B \rangle$, under the assumption that the
power spectrum is $\C_l$.  This relationship is specified by the
filter function, 
\be
\label{eq:definef}
\langle \C_B \rangle = f_{Bl} \C_l.
\ee 
Taking Eq.~\ref{eq:definef} as the definition of the filter, 
(\cite{knox99}) has shown how it can be calculated from the signal
and noise covariance matrices and the derivatives of $S$ with
respect to $\C_l$.  Taking into account all off-diagonal terms, this
prescription for $f_{Bl}$ solves problem ~\ref{nooffdiag}.  To
distinguish it from the usual practice of simply using the
window function filter, we call this the minimum-variance filter.
They are identical only in the limit of no signal correlations.

We could remove the need for filters by making the bands very narrow
since sufficiently narrow bands ensure that the sensitivity to each
$\C_l$ within the band is approximately independent of $l$.  However,
making the bands too narrow would increase the non-Gaussianity --
exasperating problem~\ref{nogauss} - because the likelihood of more
tightly constrained broad bands is better approximated by a
Gaussian. Therefore, the bands must be broad enough to have
significant constraints on their amplitudes.  For MSAM, this condition
makes the bands sufficiently broad that the sensitivity to $\cl$ varies
significantly across the bands, necessitating the use of a separate
filter for each band.

Finally, if we adopt the (\cite{bond98a}) prescription for problem~\ref{nogauss},
which requires calculation of a ``log-normal offset'', $x_B$, for
each $\C_B$, we have solutions to all four problems.  Although these solutions
are not exact, they do represent a significant improvement over the
usual flat band-power method.  

It is not necessary to break the power spectrum into bands to obtain
parameter estimates from the observations.  However, this approach
aids the comparison of different experiments with a minimum of
theoretical assumptions, as well as easing the comparison of
experimental results with theory.  By following the above procedure
for power spectrum estimation, the full weight of an experiment is
made available in an easily-tractable form for the kind of parameter
estimation outlined in Eq.~\ref{eq:chisq}

\subsection{The Application to MSAM I Data}

The MSAM1 data sets are a prime example of the limitations of the flat
band-power method.  The MSAM1 data have
high signal-to-noise and are heavily sample-variance limited when
using standard estimators of the flat band-power.
We now use the radical compression methods outlined above to probe
regions of $l$-space ignored by our previous reduction to 
flat band-powers. 

\begin{figure}[bthp]
\plotonefull{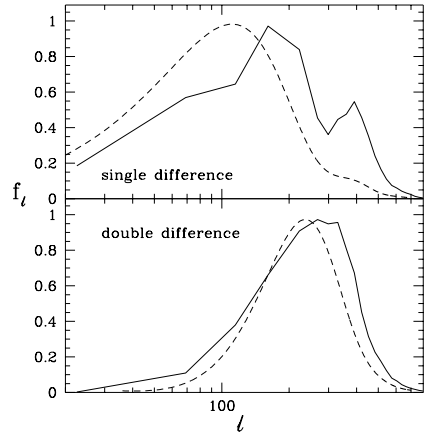}
\plotonelocal{f3.eps}
\caption[filter functions]{\baselineskip=10pt \small The minimum-variance
filters (solid lines) and window
function filters (dashed lines) for the single and double difference
demodulations.  
}
\label{fig:filts}
\end{figure}

The difference between the minimum-variance filters, $f_{Bl}$, and the
window function filters, is shown in Fig.~\ref{fig:filts}. The plot shows
the filters for each demodulation of the 3 year data for a single band
covering all $l$. Notice that the minimum-variance filters show more
response at high $l$ than the window function filters.  This is due to
the fact that the dense sampling and high signal-to-noise ratio of the
data set yield information on angular scales smaller than the
beam size.  Again, this information is in the {\it off-diagonal}
components of the covariance matrix - underscoring the need for
experiments to track the full noise covariance matrix in the data
reduction.

We plot the minimum-variance filters for the individual demodulations 
only to make the point that they are not equal to the window-function
filter as has often been assumed in the past.  In the analysis we
describe below, we do not treat the double-difference data sets
and single-difference data sets separately; the very significant
correlations between them are included. 

For this analysis, we break up the $\ell$-space coverage into three
wide bands and allow $\C_B$ to vary in each.  In line with the
discussion above, we choose the three bands such that each has enough
weight to produce an interesting constraint on the power spectrum.
The $l$-ranges for the three bands chosen are 39-130, 131-283, and 284-806.

\begin{figure}[bthp]
\plotonefull{/home/msam1/paper/msam_deltaT_map_demod.eps}
\plotonelocal{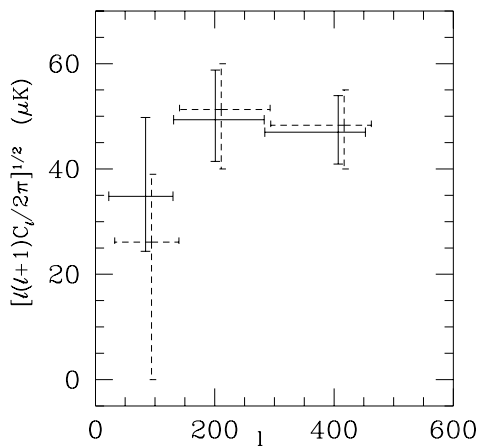}
\caption[msam bandpowers]{\baselineskip=10pt The MSAM band-power
estimates.  The solid lines
give the estimates of the power in the three bands calculated
directly from the demodulated data, and recorded in 
Table~\ref{table:pse}.  Similar results (dashed lines)
are achieved by analyzing 
a total-power map of the CMB temperature which is constructed
from the demodulated data (see section~\ref{sec:map}).
}
\label{fig:msambp}
\end{figure}

In Fig.~\ref{fig:msambp} we plot the 3 power spectrum estimates,
$\hat \C_B$ from the combined three years of data.  The central points are
located in $l$-space at $l_{\rm eff}$ which we define as the average
value of $l$ over the filter function for that band.  We include a
horizontal bar from $l_-$ to $l_+$.  For the middle band, these
are taken to be simply the beginning and end of the band ($l_- = l_< = 131$,
$l_+= l_> =283$).  For the far ends of the two outer bands, 
we take them to be where the filter falls to $e^{-1/2}$ of maximum.
Similar results (dashed lines) are achieved by analyzing 
a total-power map of the CMB temperature which is constructed
from the demodulated data.  We take the good agreement between these
two nearly independent analysis techniques as strong proof that
we have calculated the complicated signal covariances correctly.
We will discuss this procedure
in section~\ref{sec:map}.

\begin{table}[htbp]
\caption{Power Spectrum Estimates from MSAM1}
\label{table:pse}
\centering
\vspace{1ex}
\begin{tabular}{||c|c|c|c||} \hline \hline
$l_-$       & $l_{\rm eff}$    &       $l_+$   &  $\sqrt{\hat \C_B} \ (\muk)$ \\ \hline
39   & 84 & 130 & $35^{+15}_{-11}$  \\  \hline
131  & 201   & 283   & $49^{+10}_{-8}$   \\      \hline
284  & 407 & 453      & $47^{+7}_{-6}$    \\      \hline
\end{tabular}
\end{table}

While the $\hat \C_B$ are not independent, their correlation coefficients
are fairly small.  The correlation between bands 1 and 2
is $-0.18$, between bands 1 and 3 is $-0.024$ and between bands 2 and 3
is $-0.29$.  The error bars shown in Fig.~\ref{fig:msambp} 
are the result of marginalizing over the power 
in the other bands.  Under the assumption that
the other bands are fixed, the error on the band in question is 
less than 5 \% smaller.  

The power spectrum estimates, $\hat \C_B$, their weight
matrix, $F_{ll'}$, filter functions, $f_{Bl}$, as well
as log-normal offsets, $x_B$ (see (\cite{knox99})) are
available at 
\newline http://topweb.gsfc.nasa.gov  
and also
at \newline http://www.cita.utoronto.ca/~knox/radical.html
which includes similar information from other CMB 
anisotropy data sets.

\section{CMB Maps}
\label{sec:map}

A useful check of our power spectrum results can
be made by analyzing a map made from the demodulated data as 
opposed to directly from the demodulated data as we have done above.  
We begin constructing this map by recognizing
that Eq. 1 and 2 can be combined and rewritten in matrix form as
\begin{equation}
d   =  B T + n.
\end{equation}
With the assumption that the noise is Gaussian, with covariance
matrix, $N$, the most likely value of $T$, given the data, $d$,
is that which minimizes the $\chi^2$:
\begin{equation}
\chi^2 \equiv (d - B T) N^{-1} (d-B T)
.\end{equation}
This minimum, which we denote by $\hat T$, is given by
\begin{equation}
\hat T = \tilde N B N^{-1} d.
\label{eq:LINCOM}\end{equation}
This estimate of $T$ will be distributed around the
true value due to noise,  where the noise covariance matrix
is
\begin{equation}
\tilde N \equiv <(\hat T - T)(\hat T - T) > = \left( B^T N^{-1} B
        \right)^{-1}.
\label{eq:NOICOV}\end{equation}

This map can
be analyzed in the same manner as 
the demodulated data, with the
advantage that the signal covariance
matrix is now very simple to compute. Previously,
calculating the signal covariance matrix required
doing a four-dimensional integral for every covariance element.
In this new ``map basis,'' the signal covariance matrix
simplifies to
\begin{equation}
<T_i T_j > = \sum_l {2l+1\over 4\pi} P_l(\cos(\theta_{ij})) C_l
.\end{equation} The price to pay for this simplicity is that the noise
covariance, $\tilde N$, is very complicated.  We have done this
analysis as a check of the calculations in section~\ref{sec:estimate}.
The results are shown in Fig.~\ref{fig:msambp}. The agreement is a
strong argument that we have made no errors in what is a fairly
elaborate and difficult calculation.

\begin{figure}[htpb]
\plotonefull{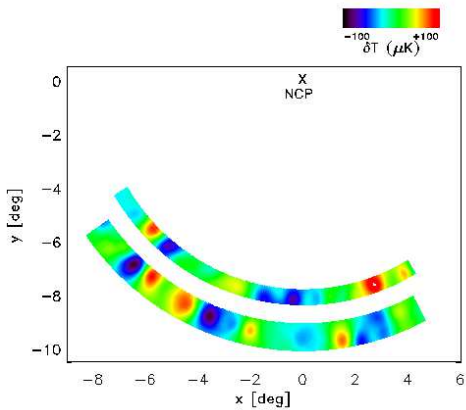}
\plotonelocal{f5.eps}
\caption{A map of the three years of data. Top region was covered by the
92/94 flight; bottom by 95 flight.}
\label{fig:map}
\end{figure}

The map, $\hat T$, is extremely noisy and not visually useful.
We can greatly reduce the noise by Wiener filtering, 
(e.g.,\cite{Bunn95,Tegmark97,knox98}).  The Wiener filter
produces the most likely $T$, given not only the data, but also an
assumed power spectrum for the signal.
The Wiener filtered maps are shown in Fig.~\ref{fig:map}.  

\section{Discussion}

Because the third band is derived by making measurements on scales
on order of the beam size, we must ask what sensitivity the
amplitude of $\hat \C_B$ has to the beam shape.  For example, if the
band sensitivity results from high frequency fluctuations in the
measurements of the various $B_i(\vec{k})$, the estimation of the
amplitude of $\C_B$ would be suspect to the errors in determining the
beammap.  We address this question by performing a number of analyses
of the three years of data.  The first analysis (leading to the quoted
values of $\hat \C_B$ here) is done using the six beammaps measured in
the three flights.  That is, MSAM92 data goes with the MSAM92
beammaps, MSAM94 data with MSAM94 beammaps, and MSAM95 data with
MSAM95 beammaps.  The second analysis is done using the beammaps
measured during the MSAM92 flight for all three years of data.  This
is followed by repeating the analysis with the beammaps from
MSAM95~\footnote{The MSAM92 and MSAM94 beammaps are similar enough to
be swapped with no change in estimated signal.}.  We find that after
accounting for the normalizations of the different data sets,
there is no evidence for the third band being sensitive
to the beammap choice. 

Reanalyzing the entire data set using beammaps
measured in flight from raster observations of Jupiter for each of the
three different years the experiment flew, is taken to be the most
pessimistic estimator of the effect of the beam on the third band. The
differences between the beammaps include all the statistical errors of
the beammaps, any errors in the raster observations themselves, and any
changes introduced by the complete rebuild, realignment, and refocusing
of the optical system and instrument configuration.

To place our estimates of the power spectrum in context,
we plot them with the predictions of several theoretical
models as well as a fit of the power in 11 bands to
all available data (based on (\cite{bond98b}) including the previously
published MSAM1 points.

\begin{figure}[bthp]
\plotonefull{/home/msam1/paper/msambp_theory_alldata.eps}
\plotonelocal{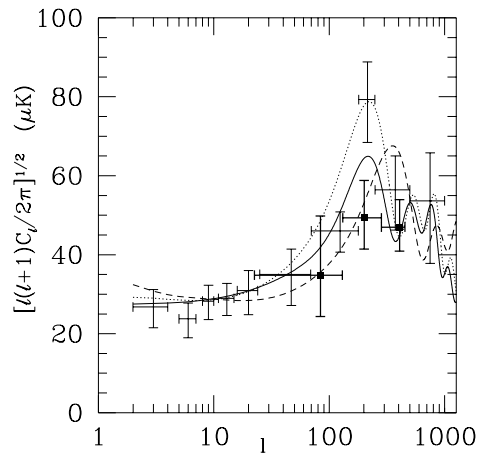}
\caption[msam bandpowers]{\baselineskip=10pt The dark error bars are
the power spectrum constraints from the 3 year MSAM data set.  The
light error bars are the result of a fit of the power in 11 bands to
all available data (based on (\cite{bond98b}) including the previously
published MSAM1 points.  The curves are standard CDM (solid), a flat
Lambda model (dotted), and an open model with $\Omega_{\rm
curvature}=0.6$ (dashed).  }
\label{fig:perspective}
\end{figure}

\section{Summary}
We have calculated new power spectrum estimates from the combined
three flights of the MSAM1 instrument.  The analysis technique
used is an improvement over the standard flat band-power aproach and includes 
all correlations in the data.  In addition to power-spectrum estimates
and their error covariance matrices we have also provided the
log-normal offsets and minimum-variance filters in order to improve
``radical compression.''
The analysis yields a strong constraint on the power spectrum at
$l\sim 400$, broadening the $l$-space coverage of the
experiment into a theoretically very interesting region.

\acknowledgements

This work would not be possible without the excellent support we
receive from the staff of the National Scientific Balloon Facility.
Financial support was provided by the NASA Office of Space Science,
under the
theme ``Structure and Evolution of the Universe.''
G.W. is supported by a National Research Administration fellowship.
K.C. is supported by a Graduate Student Researcher Program fellowship.

\bibliographystyle{aas}\bibliography{cmbr}

\end{document}